# Multi Mode (Reflection and Transmission) Operated Dielectric Resonator based Displacement Sensor

Premsai Regalla and A. V. Praveen Kumar
Department of Electrical and Electronics Engineering, Birla Institute of Technology and Science (BITS)
Pilani campus, Rajasthan-333031, India.
Premsairegalla999@gmail.com, Praveen.kumar@pilani.bits-pilani.ac.in

*Abstract*— The authors propose a multi-mode operated dielectric resonator (DR) based displacement sensor. A cylindrical dielectric resonator (CDR) is coupled to a two 50 Ω microstrip lines to enable the multi-mode operating feature. The parameters like, $|S_{11}|$ in reflection mode and $|S_{21}|$ in transmission mode is sensitive to DR displacements w.r.to transmission lines. The HFSS is carried out for numerical analysis and fabricated sensor prototype is realized with VNA measurement. The proposed sensor provides 6.2 dB/mm sensitivity over 0-4 mm range in reflection mode, and 1 dB/mm over 0-18 mm range in transmission mode. Both the HFSS and VNA responses are in good agreement to each other for this multi-mode operation. This reveals, the selection of mode freedom to the choice of operation.

*Keywords— Multi mode, Microwave resonator, Displacement sensor, Fixed frequency sensor*

## I. INTRODUCTION

Previously, RF microwave technology was limited to various applications like oscillators, amplifiers, filters, and antennas [1]. But recently, the sensor based applications are realized due to their nature of determining different physical quantities. Particularly, displacement sensors for their compact size, high sensitivity, and wide dynamic range [2]. To support this, the planar and non-planar structures use various sensing heads like different metallic and non-metallic resonators (like meta materials, co-axial, defected ground structures, frequency selective surfaces, dielectric resonators etc.,) [3]-[10]. These are generally categorized as linear (1-Dimensional, or 2-Dimensional), angular, or combined displacement sensors [3]-[10]. These sensors can be categorized into various principles like shift frequency, split in frequency, phase variance, and fixed frequency variable magnitude sensors. Among various sensing principles used, fixed frequency-variable magnitude sensors are highly robust to changes in their surrounding environment. Moreover, single frequency systems need just a low-cost, narrow-band measurement system in place of a typical costly VNA [11]-[12]. All of these sensors either operated in reflection mode or transmission mode to sense the displacement by using output variables as frequency, phase, or magnitude. Compared to displacement sensors based on other technologies (low and high frequencies), the main advantages of the microwave technologies include in small size, easy design, low fabrication costs, and high sensitivity [13].

In the existing literature, all of the displacement sensors are either operated in reflection mode or transmission mode. But the multi-mode operation i.e., combined reflection and transmission mode sensing is missing in the literature. Thus, there is a need for investigations on multi-mode operating displacement sensors which are essential in selecting mode of operation for chosen sensor application in automotive and mechanical industries. In the present paper, a DR is loaded with two microstrip lines to exhibit sensitivity to both reflection and transmission modes. The proposed resonator is operated at a single frequency. The sensor design and numerical analysis are discussed in Section. II. Experimental validation of results is presented in Section. III, followed by conclusion in Section. IV.

## II. SENSOR DESIGN AND ANALYSIS

HFSS simulated sensor schematic is presented in Fig.1. The involved design parameters are 50 Ω impedance matched microstrip lines, DR, and FR4 substrate and its specifications are mentioned in the figure description. The DR is placed exactly at the center of substrate, where its edge coincides with both microstrip lines. The DR's positional variation (*dx*) between the lines produces impedance mismatch, causing variation in the coupling level which varies both the reflection coefficient ($S_{11}$) and the transmission coefficient ($S_{21}$).

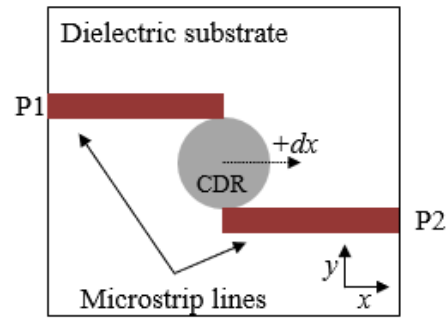

Fig.1. Schematic representation (HFSS) of proposed sensor model (**microstrip line** width =3.2 mm, **DR** diameter=19.43 mm, height= 7.3mm, **substrate** relative permittivity= 4, loss tangent= 0.02, height= 1.6 mm, size =10×10 mm$^2$)

### a) Reflection mode operation

In refection mode, the optimum impedance matching position gives the minimum $S_{11}$ at the resonant frequency of 3.87 GHz at *dx*=4 mm. The EM field distributions in DR at resonant frequency is shown in Fig. 2. The field distribution patterns help us to identify the $TE_{011+\delta}$ mode operation [14].

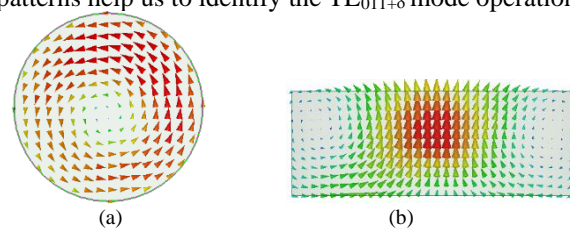

Fig.2. Simulated field distribution patterns of $TE_{011+\delta}$ mode inside the dielectric resonator at 3.87 GHz (a) E-field top-view, (b) H-field side-view

The movement of 0 to 4 mm helps to reach the optimum impedance range. Hence min $S_{11}$ at 4mm. DR further movement (>4 mm) leads to increase in $S_{11}$. The change in $|S_{11}|$ for each $dx$ helps to sense the displacement. The frequency response of $|S_{11}|$ as a function of $dx$ is shown in Fig.3. Its corresponding sensitivity curve at 3.87 GHz is shown in Fig.4. As shown, for 0-4 mm range, the proposed sensor has 6.7 dB/mm sensitivity, whereas the sensitivity is 1.2 dB/mm in the range of 4-20 mm. The above states, the reflection mode of operation supports high sensitivity with less dynamic range.

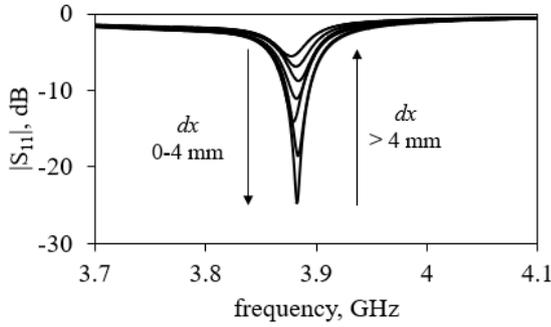

Fig.3. HFSS simulated $|S_{11}|$ spectrum of proposed sensor for each $dx$

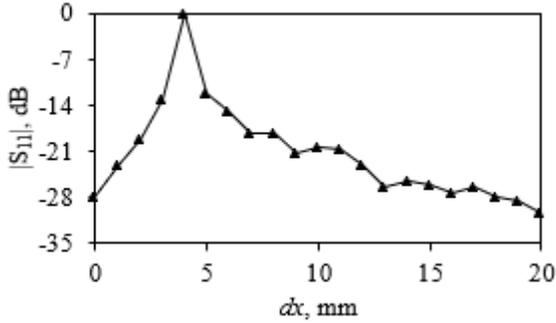

Fig.4. Sensitivity curve generated from Fig.3 at 3.87 GHz

*b) Transmission mode operation*

In transmission mode, the optimum impedance matching position gives the maximum $|S_{21}|$ at the resonant frequency of 3.87 GHz at $dx$=0 mm. The EM field distributions in DR at resonant frequency is identical to reflection mode operation as shown in Fig. 3. DR further movement (>0 mm) leads to decrease in $|S_{21}|$ helps to sense the displacement. The frequency response of $S_{21}$ as a function of $dx$ is shown in Fig.5. Its corresponding sensitivity curve at 3.87 GHz is shown in Fig.6. As shown, for each $dx$, the $|S_{21}|$ decreases monotonically at single frequency. In transmission mode, the proposed sensor shows 1.1 dB/mm sensitivity in the range of 0-19 mm. The above states, the transmission mode helps for wide dynamic range detection than reflection mode.

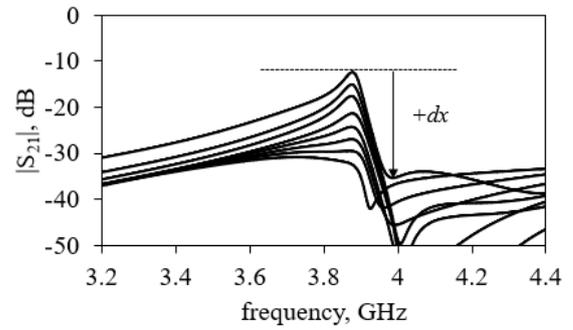

Fig.5. HFSS simulated $|S_{21}|$ spectrum of proposed sensor for each $dx$

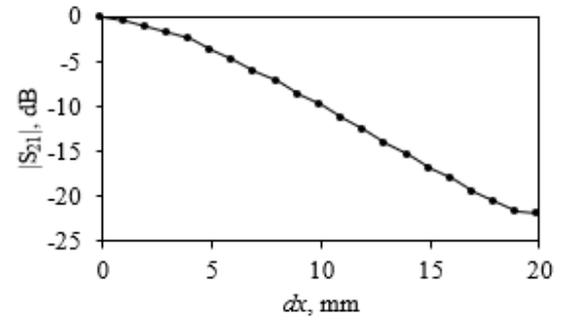

Fig.6. Sensitivity curve generated from Fig.5 at 3.87 GHz

### III. MEASUREMENT AND RESULTS

The fabricated sensor prototype is shown in Fig.7. Measurements are performed using Vector Network Analyzer (VNA) as shown in Fig. 8.

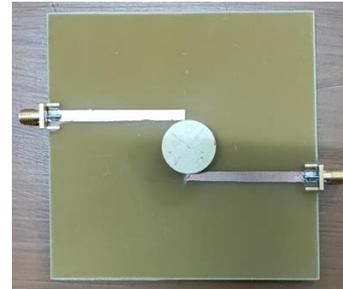

Fig.7 Fabricated prototype of the proposed sensor (dimensions are identical to Fig. 1)

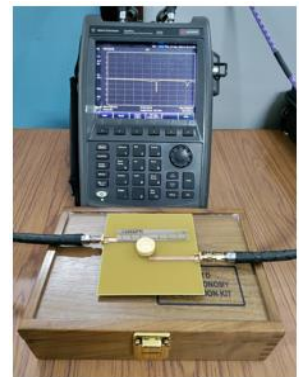

Fig.8 VNA measurement of the fabricated prototype

In reflection mode operation, the measured resonant frequency for optimum matching position at $dx = 4$ mm is 3.66 GHz while the simulated one at 3.87 GHz. Whereas, in transmission mode, the measured frequency is identical to reflection mode. The VNA measured frequency response of proposed sensor is shown in Fig.9, for both the modes. Fig.

9. (a) for reflection and Fig. 9. (b) for transmission mode. The sensitivity curves for both the reflection mode and transmission mode are shown in Fig.10 (a) and (b) respectively. The measured sensitivity curves exhibit 6.2 dB/mm in 0-4mm range in reflection mode and 1.1 dB/mm in 0-18 mm range in transmission mode operation.

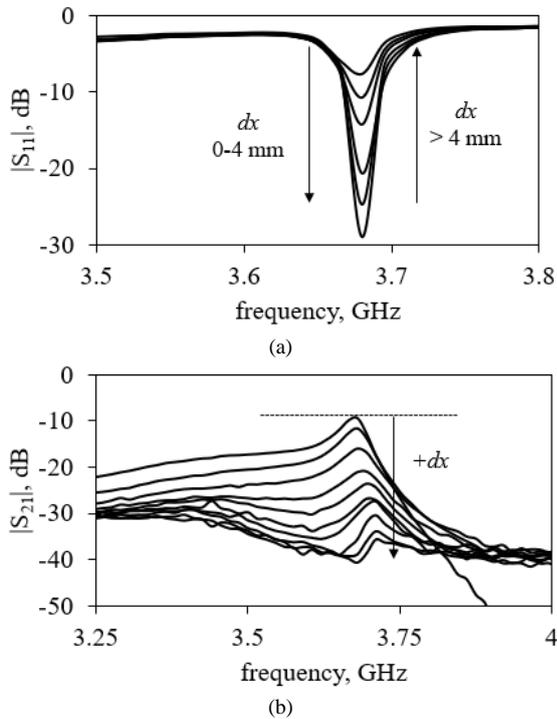

Fig.9. Measured (VNA) frequency response of the sensor for each $dx$ in (a) reflection mode and (b) transmission mode

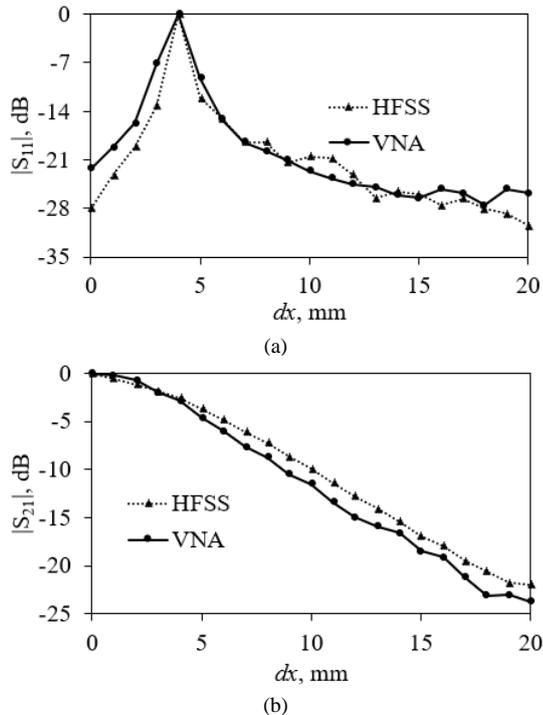

Fig.10. Comparison between the measured (VNA) and simulated (HFSS) sensitivity responses in (a) reflection mode and (b) transmission mode

## IV. CONCLUSION

The attributed sensor is capable of operating and detecting displacements in both the reflection and transmission modes with no change in the design. The sensor viability is shown with numerical and experimental results. The most important attraction of this sensor is its multi-mode single frequency operation with a choice of selectivity for chosen application like highly sensitive, or wide dynamic range. Furtherly, this single frequency sensor can easily bypass the VNA with an alternative low-cost setup for realization.